\documentclass[aps,prl,twocolumn,showpacs,amsmath,amssymb,superscriptaddress]{revtex4-1}

\usepackage{graphicx}
\usepackage{color}
\usepackage{epstopdf}
\usepackage{listings}
\usepackage{braket}

\begin{document}
\title{Super-mode spatial optical solitons in liquid crystals  with competing nonlinearities}
\author{Pawel Jung}
\affiliation{Faculty of Physics, Warsaw University of Technology, Warsaw, Poland}
\author{W. Krolikowski}
\affiliation{Laser Physics Centre, Research School of
Physics and Engineering, Australian National University, Canberra, ACT
0200, Australia}
\affiliation{Science Program, Texas A\&M University at Qatar, Doha, Qatar}
\author{Urszula A. Laudyn}
\affiliation{Faculty of Physics, Warsaw University of Technology, Warsaw, Poland}
\author{Marek Trippenbach}
\affiliation{Faculty of Physics, University of Warsaw, Warsaw, Poland}
\author{Miroslaw A. Karpierz}
\affiliation{Faculty of Physics, Warsaw University of Technology, Warsaw, Poland}

\begin{abstract}
 We study numerically formation of spatial optical solitons in nematic liquid crystals with competing nonlocal nonlinearities. We demonstrate that at the sufficiently high input power the  interplay  between focusing and thermally induced defocusing may lead to the formation of two-peak fundamental spatial solitons. These solitons have  constant spatial phase and hence represent supermodes of the self-induced extended waveguide structure. We  show  that these two-peak solitons are stable in propagation and  exhibit adiabatic transition to a single peak state under  weak absorption.

\end{abstract}
\maketitle

\section{introduction}
Spatial optical solitons represent optical beams propagating without spreading in nonlinear media due to the balance between diffraction, which tends to spread the beam and focusing from nonlinear response of the medium. They have been observed in variety of nonlinear optical materials exhibiting different types of nonlinearity including spatially local, nonlocal, Kerr-like or saturable~\cite{Stegeman:Science:99,Kivshar:book}. Typically the soliton formation has been discussed in the context of local nonlinearity, i.e. when the nonlinear response in a particular spatial location is determined by the light intensity in the very same  location. The most prominent examples are the so called  Kerr   media where the light induced refractive index change is proportional to the light intensity.   In recent decade there has been interest in the nonlocal nonlinear media, i.e. media with  the nonlinear response (index change) in a specific point  determined by the light intensity in the neighborhood of this point~\cite{Bang:pre:02}. The spatial extent of this region relative to the soliton width determines the degree of nonlocality. The nonlocal nonlinearity has been identified in  such diverse systems as thermal media~\cite{Dabby:apl:68}, nematic liquid crystals (LC)~\cite{Conti:prl:03,Assanto:book}, Bose Einstein Condensate\cite{Dalfovo:rmp:99} and  atomic clouds~\cite{Suter:pra:93,Skupin:prl:07,Maucher:prl:16}.

The typical fundamental bright optical solitons has form of  finite beam, self-trapped by the nonlinear change of the polarization  of the material. In this respect, in most situations,   the soliton is nothing but a fundamental mode of the self-induced waveguide~\cite{Snyder}.
As such, the stationary intensity profile of fundamental soliton features  single maximum. In principle it is possible to form stationary multi-peak   solitons. However, these are not fundamental solitons. They can be realised, for instance, as  vector solitons, i.e. objects formed by simultaneous propagation of few incoherently coupled optical beams (components), with  each of them  representing various (higher order) mode of the optical   waveguide induced by the total intensity.
These multi-hump solitons have been demonstrated in number of systems including photorefractive and thermal media~\cite{Mitchell:prl:98}. It is worth mentioning that in nonlinear media with spatially nonlocal nonlinearity multi-peak solitons can be formed as a bound state of two or more fundamental solitons with $\pi $ phase shift between them. In local media out-of phase solitons tend  to repel each other~\cite{Stegeman:Science:99} but strong nonlocal nonlinearity introduces attractive potential, which causes formation of bound states of solitons~\cite{Krolikowski:appa:03,Buccoliero:prl:07}.  Such dipole and higher order, multiple  solitons have been observed in nematic liquid crystals and media with thermal nonlinear response~\cite{Hutsebaut:oc:04,Rotschild:ol:06}. However,  no fundamental multi-peak  solitons have  been reported so far. While   it has been shown that model of nonlinear  media with periodic nonlocal response function  supports  multi-peak soliton, these   solitons are unstable and break up in  propagation~\cite{Esbensen:pra:12}.

In this paper we demonstrate theoretically that fundamental two-humped spatial solitons can exist in media with competing nonlocal nonlinearities~\cite{Esbensen:pra:11}. Specifically,   we discuss nonlinear model of  nematic liquid crystals. We show numerically that competition between reorientational, focusing  and thermal, defocusing  nonlinearities leads, in our configuration, to  the formation of two-peak fundamental solitons. These solitons which can be considered  as super-modes of the self-induced waveguide structure appear  to be stable and resistant to strong perturbations.   

\section{Theory and results}

We consider propagation of optical beam in the nematic liquid crystal cell comprising  LC  located  between closely placed (tens of micrometers) parallel glass plates located in the y-z plane. We assume that the internal surfaces of both plates are  conditioned (for instance, by rubbing) to ensure that the molecules of LC are  anchored and aligned at an angle $\theta=\theta_0$, with respect to the z-axis.    Hence the LC in a cell behaves  like  an uniaxial optical medium with constant refractive index. The electric field of the optical beam (wavelength $\lambda_0$ propagating in LC  changes locally the orientation of molecules, leading to the intensity-dependent index change for extraordinary polarized light. If we consider beam propagation along the z-axis, the evolution of the amplitude of electric field $E(x,z)$  is described by ~\cite{Assanto:jqe:03}
\begin{equation}\label{nls}
2ik_{o}n\left(\theta_0\right) \frac{\partial E}{\partial z} =\frac{\partial^{2} E}{\partial x^{2}}  + k_{o}^2(n^{2}\left(\theta\right)-n^{2}\left(\theta_0\right))E,
\end{equation}
where $ k_0=2\pi/\lambda_0$, and 
\begin{equation}
n\left(\theta\right) = \frac{n_{o}n_{e}}{\sqrt{n_{o}^{2}\sin^{2}\theta + n_{e}^{2}\cos^{2}\theta}} 
\label{neff}
\end{equation}
is an effective index of refraction for the y-polarized (i.e., extraordinary polarized) light.  Here $n_o$ and $n_e$ denote ordinary and extraordinary refractive indices, respectively. 
The index $n(\theta)$ depends on the local orientation of molecules, which follow the direction of electric field of the beam, here the molecular orientation angle $\theta$  is governed by the following  relation:  
\begin{equation}
\frac{\partial^{2} \theta }{\partial x^{2}} - \frac{\Delta\varepsilon \varepsilon_{o}}
{2K}\sin2\theta |E|^{2}=0. 
\end{equation}
Here $K$ is an effective elastic constant ~\cite{Assanto:jqe:03,temp:K} and $\Delta\epsilon=n_e^2-n_o^2$. As the equations Eqs.(2-3)  show, the light-induced reorientational  index change is spatially nonlocal and it  is always positive as the molecules tend to align along the direction of electric field. This leads to self-focusing of extraordinary polarized optical beam and formation of bright solitons, called nematicons~\cite{Assanto:jqe:03,nem:chiralny}.  In the following we will assume that the propagation of light in the liquid crystal is accompanied by weak absorption which causes   its heating. This process is governed by the heat equation
\begin{equation}
\kappa\frac{\partial^{2} T}{\partial x^{2}}  + \frac{c\varepsilon_0\alpha}{2} |E|^{2} = 0,
\end{equation}
where $T$ denotes temperature, $\kappa$ - thermal conductivity, $\alpha$ - absorption coefficient and $c$ is speed of light.\\
\\

\begin{figure}[htbp]
\centering
\includegraphics[width=7.50cm]{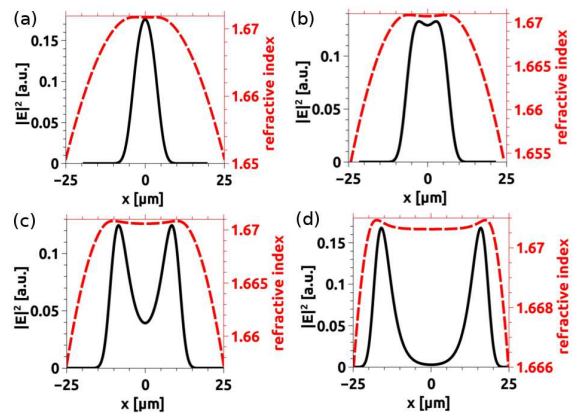}
\caption{\label{Fig1} \@ \@ \@ \@ Soliton solutions in the nematic  liquid crystal   with  competing focusing and defocusing nonlocal nonlinearities as a function of beam power. Spatial intensity (solid) and light-induced refractive index (dashed) profiles are shown (total power P= 1.4 (a), 1.8 (b), 2.0 (c) and 2.2 (d)).  }
\end{figure}

The light-induced heating  modifies both, ordinary and extraordinary  refractive indices of the liquid crystal~\cite{Li:jdt:05}, 
inducing effective  self-defocusing  of the  extraordinary polarized beam. Therefore the nonlinear response  of the nematic  liquid crystal consist of two competing, spatially nonlocal processes:  reorientation driven self-focusing and thermally-induced  self-defocusing.  
It it is worth mentioning that thermal nonlinearity alone can be also used to support bright solitons. This requires, however,  the light beam to be ordinary polarized~\cite{Derrien:joa:00,Warenghem:josab:08}.
 Here we are concerned with the role of defocusing thermal effect on reorientational nonlinearity and consider standard nematicons, formed by extraordinary polarized light. 

The temperature dependence of wide range of nematic liquid crystals is described by universal polynomial dependence~\cite{Li:jdt:05}. In this paper, for sake of concreteness   we will employ 
an empirical polynomial formula, which accurately represents thermal response of 6CHBT liquid crystal in its nematic phase,   in the temperature range 18-42 degrees (Celsius)~\cite{Dabrowski:mclc:85,Li:jap:04}. Details of this dependence are presented in the Appendix.  

In general, the elastic constant  $K$ in Eq.(3) is also  temperature-dependent~\cite{temp:K}. 
However, this dependence is week and, as we checked, has no effect on our results. Hence it will be neglected in further discussions.

The stationary bright soliton solutions of the system of equations Eqs.(1-4) we found numerically with the help of  the imaginary time method~\cite{imaginary}.  
We assume the stationary solution of Eq.(\ref{nls}) in a form: $E(x,z)=E(x)\exp(i\beta z)$, where $\beta$ is the propagation constant. Then, after taking  {\em z}  to be imaginary, $z\rightarrow iz$,  the equations Eqs.(1-4) were   solved iteratively until its solution converges to the stationary solution represent fundamental soliton of the system.  In these simulations the   temperature  and molecular orientation, where kept constant at the boundaries:  $T(x=\pm x_0)=T_0=20^\circ$  and $\theta(x=\pm x_0) =\theta_0=45^\circ$, with $2x_0$ being the width of computational window. We used finite difference to solve Eq.(3) and Eq.(4). Our simulations show that,  as long as the focusing prevails, the system always supports existence of solitons. Typical intensity profiles of these solitons  are depicted in  Fig.1, for varying total power, $P=\int_{-\infty}^{\infty}|E(x)|^2dx$.  In particular, for low input power,  when the nonlinearity is predominantly driven by molecular reorientation,
 the solitons have typical form of a single peak,  bell-like, shape [see Fig.1(a)]. However, as the power increases, the thermal effect becomes  relevant and the soliton broadens, acquiring first a flat top [Fig.1(b)] and eventually splits into two distinct peaks [Fig.1(c,d)]]. It should be stressed that the two peak 
solution still represents a fundamental soliton with  the constant phase across the soliton.   Using a waveguide analogy for solitons,  for low power, the soliton-formed  waveguide is induced purely by reorientation of molecules and is smooth,  function of spatial coordinates. When the thermal effects come to play, at higher power, the waveguide structure develops two internal peaks. This is clearly seen in refractive index profiles plotted with dashed line in Fig.1.  It is worth mentioning that although degrees of nonlocality represented by equations Eq.(3) and Eq.(4)  are comparable, they differently contribute to the  resulting refractive index. The interplay between both, orientational and thermal effects, leads  to formation of  complex waveguide structure supporting multi-peak solitons. 
\begin{figure}[htbp]
\centering
\includegraphics[width=8.0cm]{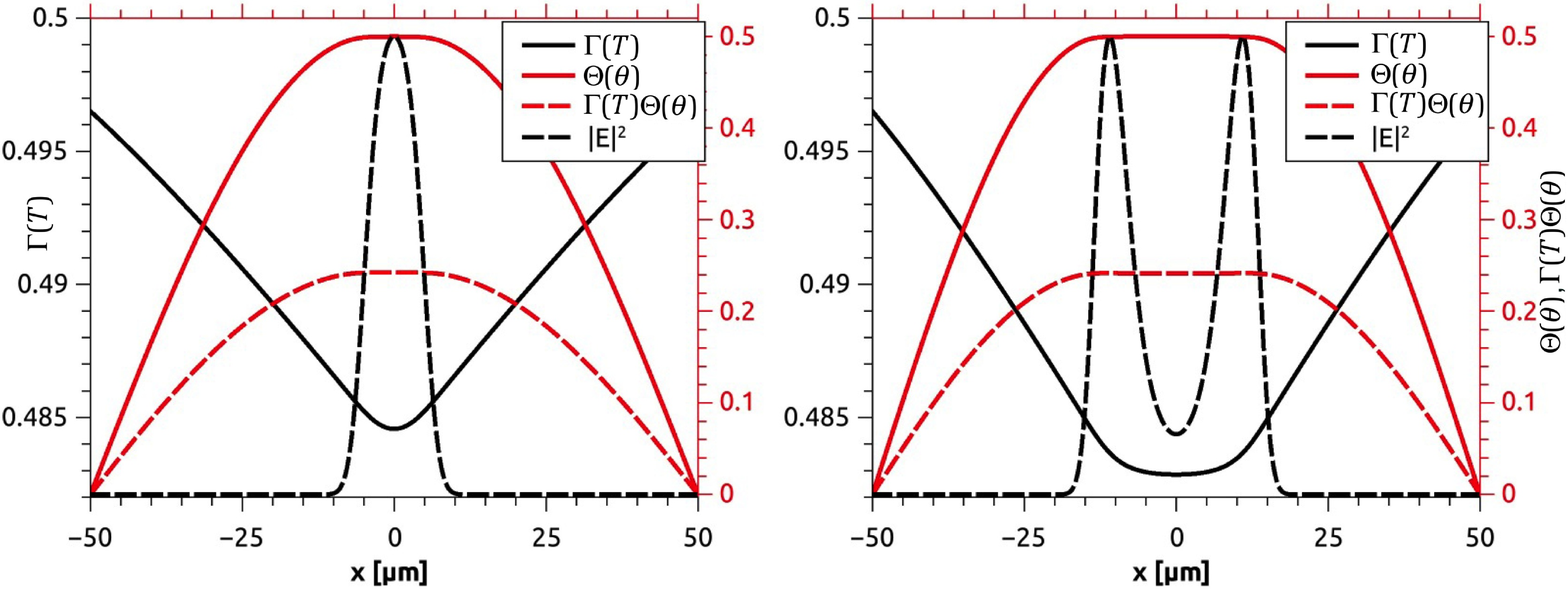}
\caption{\@ \@ \@ \@ Illustrating the interplay between reorientational ($\Theta(\theta)$ and thermal ($\Gamma(T)$)  contributions to the nonlinear response of liquid crystal induced by two different soliton intensity distributions (indicated by dashed line).  }
\end{figure}

We can get clearer picture of the competition between  thermal and reorientational mechanisms by utilizing the fact   that the anisotropy, $n_e^2-n_o^2$ , is relatively small. 
Then by  expanding  formula Eq.(1) in series we arrive at the following approximate relation for the nonlinear response 
\begin{eqnarray}
& n^2(T,\theta)-n^2(T,\theta_0) =&\nonumber\\ 
&\left (n^2_e(T)-n^2_o(T)\right)\left(\cos^2{\theta}-\cos^2{\theta_0}\right) 
=\Gamma(T)\Theta(\theta)
\end{eqnarray}
It appears that the  nonlinearity is governed  by product of two functions representing  thermal ($\Gamma(T)$) and orientational ($\Theta(\theta)$) contributions, respectively. This interplay is   different from typical nonlinear media with competing nonlinearities where different mechanisms  contribute additively to the total nonlinear response.  
We illustrate the interplay between thermal and reorientational effects in Fig.2 for two examples of light intensity distributions representing single [Fig.2(a)] and two-peak [Fig.2(b)] solitons.
  It is evident that while $\Theta(\theta)$  is responsible for   spatial  focusing due to light induced reorientation  of molecules of LC and reaches its  maximum at the intensity maximum, the 
   heat-induced contribution ($\Gamma(T)$) decreases with  intensity causing   spatial defocusing. As a result the full nonlinear response weakens and  flattens in the center, and finally develops  central dip for higher intensity. This is exactly the  regime where the two peak soliton formation takes place.   
\begin{figure}[htbp]
\centering
\includegraphics[width=7.0cm]{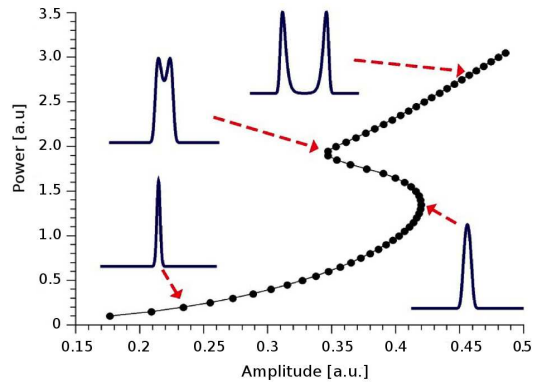}
\caption{\label{Fig2} \@ \@ \@ \@ Dependence of soliton power on its  amplitude for wavelenght $\lambda_{o}=532nm$ and initial orientation $\theta_{o}=45^{o}$, 
background temperature $T_{o}=20^{o}C$, elastic constant $K=3.6pN$, thermal conductivity $\kappa=0.135\frac{W}{m^{o}C}$ and absorption coefficient $\alpha=5.769\frac{1}{m}$. }
\end{figure}

In Fig.3 we illustrate the  relation between amplitude and the power of different soliton solutions.   It is clear that two peak solitons emerge above certain critical  power. 
The insets in Fig.3 show intensity profiles of various soliton solutions.

\begin{figure}[htbp]
\centering
\includegraphics[width=7.5cm]{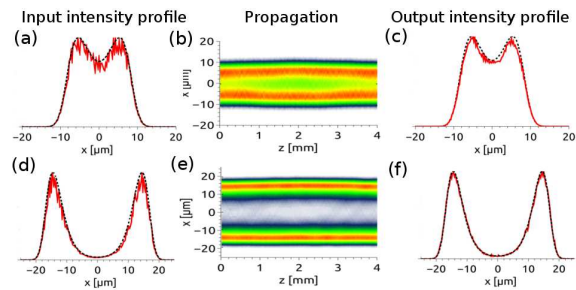}
\caption{\label{Fig3} \@ \@ \@ \@ Stability of two-peak solitons. (a, d) Input intensity  profiles. red solid (black dashed) line indicate perturbed (exact) input soliton profiles; (b,e) soliton dynamics in propagation; (c,f) final beam profiles.  }
\end{figure}

Since the bright soliton is a fundamental mode of the self-induced waveguide,  one can think of  these two-humped  solitons  as some kind of super-modes of the self-induced waveguide structure,  which are well known in the context of waveguide couplers and waveguide arrays~\cite{Hardy:jlt:85}. The nonlinear version of super-modes has been also demonstrated in  nonlinear couplers and cold atoms trapped in double well potentials \citep{cytowanie} . However, such symmetric nonlinear modes are subject to spontaneous symmetry breaking  and results in spatially asymmetric intensity distribution~\cite{Silberberg:apl:87,Matuszewski:pra:07}. Therefore it is crucial to determine the stability of our two-peak soliton solutions. To this end we used the original system of equations Eqs.(1-4) to numerically propagate soliton solution. We added random perturbation  to the amplitude of  the exact  solution and propagated it  over many diffraction lengths. We employed finite difference beam propagation method with Runge Kutta 4th order algorithm and with finite difference relaxation with multigrid algorithm technique. The results for   two-humped solitons are shown in Fig.4. The left (right) panel in each row shows the initial (final) intensity distribution (solid red line), while the dynamics of propagation is shown by the contour plots. It is clear that the solitons are stable in propagation and they retain their two-peak structure. Moreover, it is evident that intensity perturbation is smoothed out in propagation [compare left and right panel plots  in  Fig.4]. This is due to the nonlocal character of nonlinearity which tends to average out any sharp  intensity variations.  Additional (not shown) simulations confirm that at large angles solitons survive acute collisions with another solitons, however  in shallow angle collisions  their identity is lost  because the model is nonintegrable and collisions are inelastic.

\begin{figure}[htbp]
\centering
\includegraphics[width=8.5cm]{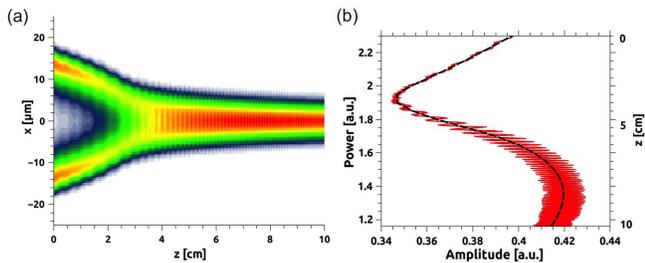}
\caption{\label{Fig4} \@ \@ \@ \@  Absorption-induced adiabatic transformation of two-humped  nematicon  into a single peak  soliton in propagation.  (a) Evolution of the beam profile. (b) Variation of the power of the soliton and its amplitude  as a function of propagation distance.  Dashed  black line  follows  stationary soliton solutions from Fig.3.}
\end{figure}

So far in our simulations we ignored linear loss. For typical experimental conditions with propagation distance of millimetres this assumption is justified. However it is interesting to see how the linear loss affects two-peak solitons in a long distance propagation. We illustrate this in Fig.5(a)  where we plot the intensity evolution  of, initially exact, two-peak soliton as its propagates in the presence of weak linear loss. As  its power decreases the solitons itself undergoes adiabatic transformation from two-peak to a single-peak structure. At any point along its propagation the beam is, in fact, a  soliton solution from the  family represented in Fig.3, as indicated by a dashed line.  This is evident in Fig.5(b) where we show evolution of power and amplitude of the beam during propagation.

Having in mind future experimental observations of the two-humped solitons we also address the issue of soliton excitation through the proper initial conditions. In Fig.6 we show numerically how the two-humped fundamental soliton can be created from the initial amplitude distribution given by two in-phase, weakly overlapping Gaussian beams. It is clear that the two-peak solitary beam is formed in propagation. The visible oscillations and emission of radiation is caused by the mismatch between exact soliton profile and input beam. Notice that unlike solitons excitation in typical nonlinear  media which is accompanied by  outward emission of dispersive waves, here  the radiation is confined to the center of the sample.  This is because  we deal here with so called infinitely nonlocal medium~\cite{Kaminer:ol:07} where  degree of nonlocality is as large as the transverse dimension  of the medium.  Consequently, the light induced waveguide is very broad extending to the sample boundaries imposing  strong  localization in the centre. 

\begin{figure}[htbp]
\centering
\includegraphics[width=8.5cm]{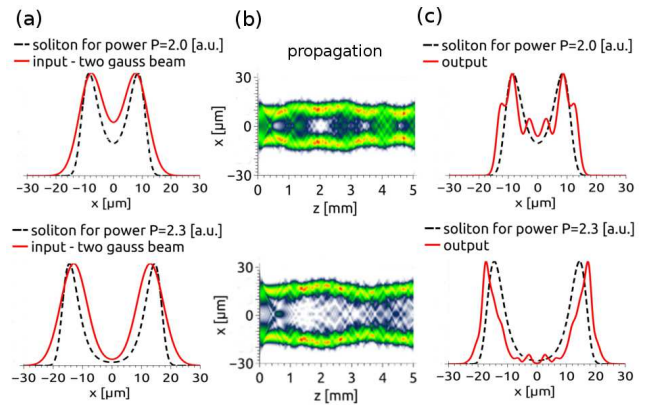}
\caption{\label{Fig4} \@ \@ \@ \@  Excitation of two-humped fundamental solitons. (a) initial amplitude distribution. (b) nonlinear evolution of the beam. (c) output intensity distribution of the soliton. The input power P=2 (top row), P=2.3 (bottom row)}
\end{figure}

\section{Conclusions}
In conclusions, we studied theoretically formation of fundamental bright solitons in nematic liquid crystals in the presence of competing nonlinearities: reorientational focusing and thermally-induced defocusing. We found that for sufficiently large input power the system supports formation of two-humped fundamental solitons with uniform phase. These solitons, which could  be considered as super-mode solitons of the self-induced two-well index structure, appear to be stable in propagation. We also showed that these solitons could be excited by properly shaped amplitude of the input beam. 
While our calculations have been conducted using parameters of a specific type of  liquid crystal, our results are applicable to a wide range of nematic liquid crystals since their  birefringence exhibits similar thermal characteristic.
\noindent
\section{Acknowledgment}
 This work was supported by the Qatar National Research Fund (grant: NPRP9-020-1-006) and the Polish National Science Centre (grant: UMO-2012/06/M/ST2/00479).
 
 \section{Appendix}
 
 The ordinary and extraordinary refractive indices of 6CHBT liquid crystal n the temperature range 18-42 degrees (Celsius)
are  modelled by the following empirical relation 
\begin{equation}
n_{o}=\Sigma_{j=0}^3 a_jT^j, \,\,\,\,\, n_e=\Sigma_{j=0}^3 b_jT^j,
\end{equation} 
where coefficients $a_j$ and $b_j$ are also function of the wavelength. 
For instance,  for  $\lambda$=532~nm,  the coefficients $a_j$ and $b_j$ read:
\begin{eqnarray}
a_0&=&\!1.659,\, a_1=2.814\times \!10^{-3} ,\,a_2=-0.103\times \!10^{-3},\nonumber \\
b_0&=&\!1.545, \, b_1=-1.861\times \!10^{-3} ,\, b_2=3.118\times \!10^{-5}. \nonumber \\
\nonumber
\end{eqnarray}
where T is expressed in degrees (Celsius).

Figure Fig.7 illustrates the  above temperature dependence.  It clearly shows that while ordinary refractive index only weakly depends on  temperature, the  extraordinary 
index decrease fast with temperature. At roughly 42$^\circ$ the crystal undergoes  phase transition  from nematic to isotropic phase.  
\begin{figure}[htbp]
\centering
\includegraphics[width=7.5cm]{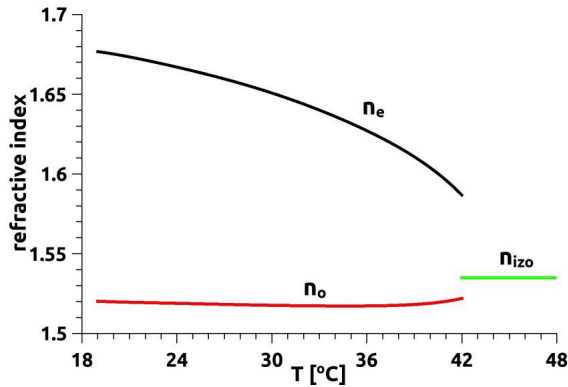}
\caption{\label{Fig4} \@ \@ \@ \@  Illustrating temperature dependence of ordinary ($n_o$) and extraordinary ($n_e$) refractive indices of 6CHBT liquid crystal,  for $\lambda$=532nm. $n_{izo} $ denotes the refractive index in the isotropic phase.}
\end{figure}



\end{document}